\documentclass[12pt,onecolumn,draftcls]{IEEEtran}
\usepackage[dvips]{graphicx}
\usepackage{amscd}
\usepackage{amsmath,epsfig}
\usepackage{amssymb}
\usepackage{amsmath}
\usepackage{cases} 
\usepackage{amssymb,amsmath,cite,color}
\usepackage{epsfig}
\usepackage{slashbox}
\usepackage{bm}
\usepackage{color}

\newcommand{\be}{\mathbf{e}}

\newcommand{\bp}{\mathbf{p}}
\newcommand{\bq}{\mathbf{q}}
\newcommand{\bc}{\mathbf{c}}

\newcommand{\bx}{\mathbf{x}}

\newcommand{\bA}{\mathbf{A}}

\newcommand{\tabincell}[2]{\begin{tabular}{@{}#1@{}}#2\end{tabular}}

\renewcommand{\frac}{\dfrac}

\newcommand{\SI}{{\mbox{SINR}}}

\newcommand{\K}{{\cal K}}
\newcommand{\N}{{\cal N}}

\newtheorem{dingyi}{Definition~}
\newtheorem{dingli}{Theorem~}

\begin{document}

\title{\huge Joint Power and Admission Control based on Channel Distribution Information: A Novel {Two-Timescale} Approach
\author{Qitian Chen, Dong Kang, Yichu He, Tsung-Hui Chang, and Ya-Feng Liu}
\thanks{Copyright (c) 2015 IEEE. Personal use of this material is permitted. However, permission to use this material for any other purposes must be obtained from the IEEE by sending a request to pubs-permissions@ieee.org.}
\thanks{Q. Chen and Y. He are with the School of Mathematical Sciences, Peking University, Beijing 100871, China (\{chenqitian,heyichu\}@pku.edu.cn).}
 \thanks{D. Kang and Y.-F.~Liu (corresponding author) are with the State Key Laboratory
of Scientific and Engineering Computing, Institute of Computational
Mathematics and Scientific/Engineering Computing, Academy of
Mathematics and Systems Science, Chinese Academy of Sciences,
Beijing 100190, China (e-mail: \{kangdong,yafliu\}@lsec.cc.ac.cn). The work of Y.-F. Liu is supported in part by the National Natural
Science Foundation of China under Grants 11671419, 11631013, 11331012, and 11571221.}
\thanks{T.-H. Chang is with the School of Science and Engineering, The Chinese
University of Hong Kong, Shenzhen, and the Shenzhen Research Institute of Big Data, Shenzhen 518172, China (e-mail: tsunghui.chang@ieee.org). The work of T.-H. Chang is supported in part by the National Natural Science Foundation of China under Grant 61571385.}
\thanks{The authors thank Professor Zaiwen Wen of Peking University for many useful discussions on an early version of this paper.}}

 \maketitle \begin{abstract}
     \boldmath
In this letter, we consider the joint power and admission control (JPAC) problem by assuming that only the channel distribution information (CDI) is available. Under this assumption, we formulate a new chance (probabilistic) constrained JPAC problem, where the signal to interference plus noise ratio (SINR) outage probability of the supported links is enforced to be not greater than a prespecified tolerance. To efficiently deal with the chance SINR constraint, we employ the sample approximation method to convert them into finitely many linear constraints. Then, we propose a convex approximation based deflation algorithm for solving the sample approximation JPAC problem. Compared to the existing works, this letter proposes a novel {two-timescale} JPAC approach, where admission control is performed by the proposed deflation algorithm based on the CDI in a large timescale and transmission power is adapted instantly with fast fadings in a small timescale. The effectiveness of the proposed algorithm is illustrated by simulations.
     \end{abstract}

\section{Introduction}

JPAC has been recognized as an effective interference
management scheme in a variety of wireless scenarios \cite{convex_approximation,region,removals,Simple,ex3,msp,Lq,ACM,scheduling,powerx,bbound,ex1,ex2,performance,distributed,robust,decentralized,tsung-hui,SPAWC,q-norm}, including cellular, ad hoc, and underlay cognitive networks. The goal of JPAC is to support a maximum number of links to achieve specified SINR targets with a minimum total transmission power. 
The JPAC problem is NP-hard in general \cite{convex_approximation,removals,msp}, so various heuristic algorithms have been proposed \cite{convex_approximation,region,removals,Simple,ex3,msp,Lq,ACM,scheduling,powerx,bbound,ex1,ex2,performance,distributed,robust,decentralized,tsung-hui,SPAWC,q-norm}. Most of the existing works on JPAC assume that the transmitters have perfect instantaneous channel state information (CSI) of the receivers. For instance, under the perfect CSI assumption, \cite{convex_approximation} and \cite{msp} proposed linear programming based deflation algorithms for solving the JPAC problem. 

However, the assumption of perfect CSI generally does not hold true in practice due to CSI estimation errors or limited CSI feedback \cite{Outage12,feedback}. Even though instantaneous CSI can be made perfectly known, performing admission control according to the fast time-varying channels would lead to excessively high computational and signaling costs. In view of this, \cite{robust} formulated the JPAC problem as a chance SINR constrained program by assuming that only the CDI is available. {Compared to the CSI, the CDI can remain unchanged over a relatively long period of time. Therefore, admission control based on the CDI can be performed in a larger timescale, thus reducing the computational cost and signaling overhead significantly.  

In this letter, we propose a new chance SINR constrained JPAC formulation under the CDI assumption and also a novel {two-timescale} approach to solving it. In our proposed approach, admission control is performed based on the CDI in a large timescale and power control is done in accordance with the CSI (but without knowing the CSI) in a small timescale. The key difference between the proposed new formulation/approach and that in \cite{robust} lies in that the proposed one allows the transmitters to adapt transmission powers with instantaneous channel fadings, whereas the one in \cite{robust} used constant powers. Such a two-timescale approach is desirable in practice, as power adaption usually costs little signaling overhead while can significantly improve the system performance (for instance, significantly reduce the total transmission power). To solve the chance constrained JPAC problem (i.e., perform admission control in a large timescale), the sample approximation scheme \cite{distributionally,uncertainty} is used to convert the chance SINR constraint into finitely many linear constraints. We show (in Theorem \ref{thm-sample}) that the chance constraint can be well approximated by its sample approximation as long as the sample size is sufficiently large. We further reformulate the sample approximation JPAC problem as a group sparse minimization problem and propose a convex approximation based deflation algorithm (see Algorithm 1 further ahead) for solving it. Numerical simulations show that our proposed algorithm performs much better than the algorithm in \cite{robust}.


We adopt the following notations in this letter. We denote the index sets $\left\{1,2,\ldots,K\right\}$ and $\left\{1,2,\ldots,N\right\}$ by $\K$ and $\N,$ respectively. We use $\left|\cal S\right|$ to denote the cardinality of set $\cal S.$ We use $\mathbb{E}(\cdot)$ to denote the expectation operator. For a given vector $\bx,$ $\|\bx\|_0$ stands for its indicator function (i.e., $\|\bx\|_0=0$ if $\bx=\mathbf{0}$ and $\|\bx\|_0=1$ otherwise). All stacking of matrices, vectors, and scalers will be written in \textsc{Matlab} language.
Finally, we use $\be$ and $\mathbf{0}$ to represent the all-one and all-zero vectors of an appropriate size, respectively.



\section{System Model and Review}

In this section, we introduce the system model and briefly review the related work \cite{robust}.

\subsection{System Model} We consider a $K$-link (transmitter-receiver pair) single-input single-output interference channel with channel gains $g_{k,j}\geq 0$ (from transmitter $j$ to receiver $k$), noise power $\eta_k>0$, SINR target $\gamma_k>0$, and power budget $\bar{p}_k>0$ for all $k,j\in \mathcal{K}$. Assume that channel gains $\left\{g_{k,j}\right\}$ are random variables defined in the probability space $(\Omega,\mathcal{F},\mathbb{P})$ and there is a central controller who knows the channel distribution. To highlight the dependence of $\{g_{k,j}\}$ on $\Omega,$ we write $\{g_{k,j}\}$ as $\{g_{k,j}^{\omega}\}$ in this letter. 

\subsection{Brief Review of \cite{robust}}
To streamline the presentation and to make the contribution of this letter more clear, we briefly review a closely related work \cite{robust}. In \cite{robust}, {the JPAC problem based on the CDI concerns supporting as many as possible links satisfying specified chance SINR requirement with a minimum total transmission power.}
Specifically, let $p_k$ be the transmission power of link $k.$ In \cite{robust}, a set $\cal S$ of links is defined as supported if there exist $\left\{p_k\right\}$ such that \begin{equation}\label{chance}\mathbb{P}\left(\frac{g_{k,k}^{\omega}p_k}{\eta_k+\sum\limits_{j\neq k}g_{k,j}^{\omega}p_j}\geq \gamma_k,\,k\in{\cal S}\right)\geq 1-\epsilon,\end{equation}
which implies that the SINR outage probability of all links in $\cal S$ should be not larger than a specified tolerance $\epsilon\in(0,1).$
It was proposed in \cite{robust} to use $N$ of SINR samples, i.e., \begin{equation}\label{sample}
  \frac{g_{k,k}^np_k}{\eta_k+\displaystyle\sum_{j\neq
k}g_{k,j}^np_j}\geq \gamma_k,~k\in{\cal S},~n\in{\N},
\end{equation} to approximate \eqref{chance}, where $\{g_{k,j}^n\}_{n=1}^N$ are independent samples drawn according to the distribution of $\{g_{k,j}^{\omega}\}.$ It was shown in \cite{distributionally,uncertainty} that, if the sample size $N$ is not less than 
\begin{equation}\label{N-K*}{N_1^*:=\left\lceil\frac{1}{\epsilon}\left(K-1+\ln\frac{1}{\delta}+\sqrt{2(K-1)\ln\frac{1}{\delta}+\ln^2\frac{1}{\delta}}\right)\right\rceil}\end{equation} for any $\delta\in(0,1),$ then any solution to \eqref{sample} will satisfy \eqref{chance} with probability at least $1-\delta.$

Furthermore, \cite{robust} developed a deflation algorithm for solving the sample approximation JPAC problem, which maximizes the number of supported links in the sense of \eqref{sample} and at the same time minimizes the total transmission power.


\section{Proposed Chance SINR Constrained JPAC Formulation}
In this section, we present our new chance constrained JPAC formulation. We first give a new definition for the supported set of links.
\begin{dingyi}
   For any $\epsilon\in(0,1),$ a set $\cal S$ of links is supported if there exist $\left\{p_k^{\omega}\right\}$ such that \begin{equation}\label{chance2}{\mathbb{P}\left(\frac{g_{k,k}^{\omega}p_k^{\omega}}{\eta_k+\sum\limits_{j\neq k}g_{k,j}^{\omega}p_j^{\omega}}\geq \gamma_k,\,k\in\cal S\right)\geq 1-\epsilon,}\end{equation}
\end{dingyi} where the transmission power $p_k^\omega$ (of link $k$) is a nonnegative function defined over the sample space $\Omega.$

It can be seen that the key difference between \eqref{chance} and \eqref{chance2} is that the transmission power is set to be a constant number over the sample space in \eqref{chance} whereas the transmission power is a function defined over the sample space in \eqref{chance2}. This implies that the transmission power $\left\{p_k^{\omega}\right\}$ in \eqref{chance2} is adaptive with channel fadings. Therefore, the proposed JPAC approach based on \eqref{chance2} is expected to perform better than that based on \eqref{chance}. 
 %

%
%
}


The following Theorem \ref{thm1} presents {the new mathematical formulation} of the chance SINR constrained JPAC problem. 
Theorem \ref{thm1} can be proved similarly as in \cite{msp} and \cite{robust} and a detailed proof is provided in Section I of \cite{companion}.
\begin{dingli}\label{thm1}Suppose the parameter $\alpha$ satisfies
\begin{equation}\label{alpha1}0< \alpha<\alpha_1:=1/{\be^T\bar\bp}.\end{equation}
Then the JPAC problem can be formulated as
\begin{equation}\label{model}
\begin{array}{cl}
\displaystyle \max_{\mathcal{S},\,\mathbf{p}^{\omega}}&|\mathcal{S}|-\alpha \displaystyle\be^T\mathbb{E}({\bp^{\omega}})\\
\mbox{\text{s.t.}}
  &\eqref{chance2}~\text{and}~\mathbf{0} \leq \bp^{\omega}\leq \bar{\bp},\forall~\omega\in\Omega,
  \end{array}
\end{equation}
where $\mathcal{S}$ denotes the set of supported links, 
$\bp^\omega=(p_1^\omega,p_2^\omega,\ldots,p_K^\omega)^T$ denotes the transmission power of all links at the point $\omega\in\Omega,$ and $\bar
\bp=(\bar p_1,\bar p_2,\ldots,\bar p_K)^T$ denotes the power budget of all links.
\end{dingli}

In the objective of \eqref{model}, the first term maximizes the number of supported links, the second term minimizes the total expected transmission power, and the parameter $\alpha$ is to balance the two terms. {In particular, formulation \eqref{model} with $\alpha$ satisfying \eqref{alpha1} is capable of finding the maximum admissible set with a minimum total transmission power (from potentially multiple maximum admissible sets).}  



Similar in \cite{robust}, we propose to handle the difficult chance SINR constraint \eqref{chance2} via the sample approximation. Suppose that $\{g_{k,j}^n\}_{n=1}^N$ and $\{p_k^n\}_{n=1}^N$ are independent samples drawn according to the distribution of $\{g_{k,j}^{\omega}\}$ and $\left\{p_k^{\omega}\right\},$ we use 
\begin{eqnarray}\label{samap}
\frac{g^n_{k,k}p^n_k}{\eta_k+\sum\limits_{j\neq k }g^n_{k,j}p^n_j}\geq\gamma_k,~k\in{\cal S},~n\in\mathcal{N}
\end{eqnarray}
to approximate (\ref{chance2}). Notice that \eqref{samap} is different from \eqref{sample} in the sense that the power in \eqref{samap} is adaptive with $\{g^n_{k,j}\}.$

We want to infer whether \eqref{chance2} is true based on its sample approximation \eqref{samap}.
Unfortunately, the analysis results in \cite{distributionally,uncertainty} are not applicable to our case because they deal with the case where the number of design variables (and design variables themselves) does not change with the sample size (the samples) while the number of design variables $\{p_k^n\}$ in our case increases (linearly) with the sample size. To overcome this, we leverage the statistical hypothesis testing theory \cite{hypothesis}.~
The following theorem shows that the sample approximation performance can be guaranteed if the sample size is sufficiently large.



\begin{dingli}\label{thm-sample}
  For any $\epsilon \in (0,1)$ and $\delta \in (0,1),$ suppose that the sample size $N$ in \eqref{samap} is greater than or equal to
\begin{eqnarray}\label{N}
N_2^*:=\left\lceil \frac{-2}{\epsilon^2\ln{{\delta}}} \right\rceil.
\end{eqnarray} For any given set ${\cal S}\subseteq \K$, if \eqref{samap} is feasible, then \eqref{chance2} is satisfied with a significance level $\delta$.
\end{dingli}

The proof of Theorem \ref{thm-sample} can be found in Section II of \cite{companion}. 
Note that although $N_2^*$ is of order ${\cal O}\left(1/\epsilon^2\right)$ while $N_1^*$ is of order ${\cal O}\left(1/\epsilon\right),$ $N_2^*$ does not depend on $K$ as $N_1^*$ does. 



Moreover, by using the sample average to approximate the expectation, we obtain the following sample approximation JPAC formulation
\begin{equation}\label{model-sample}
\begin{array}{cl}
\displaystyle\max_{\mathcal{S},\left\{p_k^n\right\}}&|\mathcal{S}|-\displaystyle
\frac{\alpha}{N}\sum_{k\in\K}\sum_{n\in\N}p_k^n\\[15pt]
\mbox{s.t.} 
  &\eqref{samap}~\text{and}~0 \leq p_k^n \leq {{\bar p}}_k,~k\in{\K},~n\in \mathcal{N}.
\end{array}\end{equation}

Two remarks on problem \eqref{model-sample} are in order. First, $\{g_{k,j}^n\}_{n=1}^N$ in \eqref{model-sample} are \emph{not} instantaneous channel gains; they are independent samples generated according to the distribution of $\{g_{k,j}^{\omega}\}$. Second, as $\{p_k^n\}$ are optimized with $\{g_{k,j}^n\}$ in contrast to $\{p_k\}$ in [16], problem \eqref{model-sample} is expected to yield better performance than the sample approximation in [16].



For simplicity, in the sequel we will refer a set $\cal S$ of links to be supported if all the sample approximation constraints in \eqref{samap} are satisfied .

\section{A Novel {Two-Timescale} JPAC Approach}
In this section, we develop a convex approximation based deflation algorithm for solving problem \eqref{model-sample} based on a group sparse minimization reformulation. Then, we propose a novel {two-timescale} JPAC approach by combining the proposed deflation algorithm with the Foschini-Miljanic algorithm \cite{Simple}.


\subsection{Group Sparse Minimization Reformulation}
For ease of presentation, let us define
\begin{eqnarray*}
\mathbf{c}_k=\left(c_k^1,c_k^2,\ldots,c_k^N\right)^{T}\in
\mathbb{R}^{N\times1},~k\in\K,
\end{eqnarray*} and
\begin{eqnarray*}
{{\bA}_k=
\begin{pmatrix}
\left(\mathbf{a}_{k}^1\right)^T& \mathbf{0} &\mathbf{0}&\mathbf{0}\\
\mathbf{0}& \left(\mathbf{a}_{k}^2\right)^T&\mathbf{0}&\mathbf{0}\\
\mathbf{0}& \mathbf{0}& \ddots& \mathbf{0}\\
\mathbf{0}& \mathbf{0}&\mathbf{0} &\left(\mathbf{a}_{k}^N\right)^T\\
\end{pmatrix}
\in \mathbb{R}^{N\times NK},~k\in\mathcal{K},}\\
\end{eqnarray*}where $$c_k^n=\frac{\gamma_k\eta_k}{g_{k,k}^{n}\bar{p}_k},~n\in{\N},~k\in{\K},$$\vspace{-0.2cm}$$\mathbf{a}_k^n=\left(a_{k,1}^n,a_{k,2}^n,\ldots,a_{k,K}^n\right)^T\in\mathbb{R}^{K\times 1},~n\in{\N},~k\in{\K},$$\vspace{-0.2cm}
\begin{equation*}\label{A}
a_{k,j}^n=\left\{\begin{array}{cl}
1,&\text{if~}k=j;\\[3pt]
\displaystyle - \frac{\gamma_kg_{k,j}^n\bar p_j}{g_{k,k}^n\bar p_k},&\text{if~}k\neq j.
\end{array}
\right.
\end{equation*} Note that $\{a_{k,j}^n\}$ are normalized sampled channel gains. Besides, let 
${\bq}=\left({\bq}^1;\ldots;{\bq}^N\right)\in\mathbb{R}^{NK\times1},$
   where $${\bq}^n=\left(\frac{p_1^n}{\bar{p}_1},\frac{p_2^n}{\bar{p}_2},\ldots,\frac{p_K^n}{\bar{p}_K}\right)^{T}\in\mathbb{R}^{K\times1},~n\in\mathcal{N}.$$

With the above notation, it can be checked that $\SI_k^n\geq \gamma_k$ if and only if $\left(\mathbf{a}_k^n\right)^T\bq^n\geq c_k^n.$ Therefore, all constraints in \eqref{samap} are satisfied if and only if $\bA_k\bq\geq\bc_k$ for all $k\in\cal S.$ It follows from this fact, the definition of $\|\cdot\|_0,$ and the balancing lemma \cite[Proposition 1]{performance} that problem \eqref{model-sample} is equivalent to the following group sparse optimization problem
\begin{equation}\label{L0}\begin{array}{rl}
\displaystyle \min_{\bq}&\displaystyle\sum_{k\in\K} \left\|{\bA}_k {\bq}-\bc_k\right\|_0+\displaystyle
\frac{\alpha}{N}\sum_{n\in\N} \bar{\mathbf{p}}^T\mathbf{q}^n\\[8pt]
\displaystyle \mbox{s.t.}&\mathbf{0}\leq\mathbf{q}^n\leq\mathbf{e},~n\in{\N}.
\end{array}\end{equation}

\subsection{A Convex Approximation based Deflation Algorithm}
Although problem \eqref{L0} is still hard to solve (due to discontinuous $\|\cdot\|_0$), it allows for an efficient convex approximation. We use the mixed $\ell_2/\ell_1$ norm \cite{group} to approximate the $\ell_0$ norm in \eqref{L0} and obtain the following convex approximation problem
\begin{equation}\label{L2}\begin{array}{rl}
\displaystyle \min_{\bq}&\displaystyle\sum_{k\in\K} \left\|{\bA}_k {\bq}-\bc_k\right\|_2+\displaystyle
\frac{\alpha}{N}\sum_{n\in\N} \bar{\mathbf{p}}^T\mathbf{q}^n\\
\displaystyle \mbox{s.t.}&\mathbf{0}\leq\mathbf{q}^n\leq\mathbf{e},~n\in{\N},
\end{array}\end{equation}which can be {reformulated as a second-order cone program by introducing auxiliary variables} and thus solved efficiently by CVX \cite{cvx}.


The solution $\bar \bq$ to problem \eqref{L2} can be used to check the simultaneous supportability of all links. If all links cannot be simultaneously supported, we propose to remove the link with the largest interference plus noise footprint \cite{removals}:
\begin{equation}\label{removal}
k_0=\arg\displaystyle\max_{k\in\K}\left\{ \sum\limits_{j\neq k} |a_{k,j}^{\bar{n}_k}| \bar{q}_j^{\bar{n}_k} +\sum\limits_{j\neq k}
|a_{j,k}^{\bar{n}_j}|\bar{q}_k^{\bar{n}_j}+\eta_k\right\},
\end{equation}
where $\bar{n}_k=\arg\displaystyle\max_{n\in\N}\{{c}_k^n-\left(\mathbf{a}_k^n\right)^T\bar{\mathbf{q}}^n\}.$ 

%
%
%

The proposed convex approximation based deflation algorithm for solving problem \eqref{model-sample} is given as Algorithm 1. The key difference between the proposed algorithm and the one in \cite{robust} lies in \textbf{Step 2}, albeit the framework of them is the same.
{Algorithm 1 is of polynomial time complexity, i.e., it has a complexity of ${\cal O}\left(K^{4.5}\right)$, since it
needs to solve at most $K$ of problems \eqref{L2} and solving one problem in the form of \eqref{L2} requires ${\cal O}\left(K^{3.5}\right)$ arithmetic operations \cite[Page 423]{com3}. In contrast, the computational complexity of the SOCP-D algorithm in \cite{robust} is ${\cal O}\left(K^{8}\right).$} The postprocessing step (\textbf{Step 3}) aims at admitting the links
removed in the admission control step as in \cite{robust}.

\begin{center}
\framebox{
\begin{minipage}{12.5cm}
\flushright
\begin{minipage}{12.5cm}
\centerline{\bf Algorithm 1: A Convex Approximation }
\centerline{\bf based Deflation Algorithm for Solving Problem \eqref{model-sample}}
\vspace{0.05cm} \textbf{Step 1.} Initialization: Input
data
$\left\{\bA_k,\bc_k,\bar p_k\right\}$ and $c\in(0,1).$\\[2.5pt]
\textbf{Step 2.} Admission control: Solve problem \eqref{L2} with $\alpha=c\alpha_1,$ where $\alpha_1$ is defined in \eqref{alpha1}; check whether all
links are supported: if yes, go to \textbf{Step 3}; else remove link $k_0$ according to
\eqref{removal}, set ${\K}={\K}\setminus\left\{k_0\right\},$ and repeat \textbf{Step 2}.\\[2.5pt]
%
%
\textbf{Step 3.} Postprocessing: Check the removed links for
possible admission.
\end{minipage}
\end{minipage}
}
\end{center}



\subsection{A Novel {Two-Timescale} JPAC Approach}

As the proposed deflation Algorithm 1 requires only the CDI, it can be implemented in a large timescale. More specifically, the central controller first generates the channel samples following the CDI, runs Algorithm 1 to determine the set ${\cal S}_0$ of supported links, and then inform it to all links. \emph{Since the transmission power $\{p_k^n\}$ in \eqref{model-sample} are based on the channel samples, they cannot be used for real-time power control.} {However,
the links in set ${\cal S}_0$ can adapt their transmission power based on instantaneous CSI. 
%
%
%
%
%
Interestingly, by employing the Foschini-Miljanic algorithm \cite{Simple}, the transmitters can do power control efficiently and distributively without the need of knowing CSI (the algorithm relies on the receivers iteratively feeding back their SINR to the transmitters).
}


In summary, in the above JPAC approach---a smart combination of proposed Algorithm 1 and the Foschini-Miljanic algorithm \cite{Simple}, admission control is done by the central controller in a relatively large timescale based on the CDI while power control is performed by supported links in a small timescale. Therefore, our proposed JPAC approach is a {two-timescale} approach, which is not only robust in the sense that it supports a fixed set of links over a relatively long period of time but also power efficient since power control is adaptively performed based on instantaneous channel conditions.


\section{Numerical Simulations}\label{sec:simulation}
In this section, we present numerical simulation results to illustrate the effectiveness of our proposed Algorithm 1. We use the (average) number of supported links and the (average) total transmission power as the metrics for {comparing} our proposed algorithm with the PABB-D algorithm in \cite{robust}, {which is the only algorithm (based on our knowledge) that addresses the JPAC problem based on the CDI without specifying any particular distribution}.  We also adopt the NLPD algorithm in \cite{msp} as our benchmark, {which provides an upper bound on the number of supported links, albeit the upper bound is generally not achievable by our proposed algorithm (because the NLPD algorithm requires perfect CSI).} 

Our simulation setup is the same as the ones in \cite{convex_approximation,robust}, i.e., each transmitter's location obeys the uniform distribution over a {$2$ Km $\times$ $2$ Km} square and the location of each receiver is uniformly generated in a disc with center at its corresponding transmitter and radius $400$ m, excluding a radius of $10$ m. Suppose that $\{g_{k,j}^{\omega}\}$ are generated from the Racian channel model \cite{goldsmith}, i.e.,
\begin{equation}\label{gains}
{g_{k,j}^{\omega}=\left|\sqrt{\frac{\kappa}{\kappa+1}}+\sqrt{\frac{1}{\kappa+1}}\zeta^{\omega}\right|^2\frac{1}{d_{k,j}^4},~k,j\in\mathcal{K},}
\end{equation}
%
where $\zeta^{\omega}$ obeys the standard Gaussian distribution, $d_{k,j}$ is the Euclidean distance from transmitter $j$ to receiver $k$, and $\kappa$ is the ratio of the power in the line of sight component to the power in the other multipath components. Each link's SINR target is set to be $\gamma_k=2$ dB, noise power is set to be $\eta_k=-90$ dB, and power budget is set to be $\bar{p}_k=3\underline{p}_k,$ where $\underline{p}_k$ is the minimum power for link $k$ to satisfy its SINR target without any interference from others when $\kappa=+\infty.$ {We set $(\epsilon, \delta)$ in \eqref{N} to be $(0.05,0.01);$ set $c$ in the proposed Algorithm 1 to be $0.999;$ and set $\kappa$ in \eqref{gains} to be $100.$}


%

%

\begin{figure}[!t]
\centering\includegraphics[width=8.5cm]{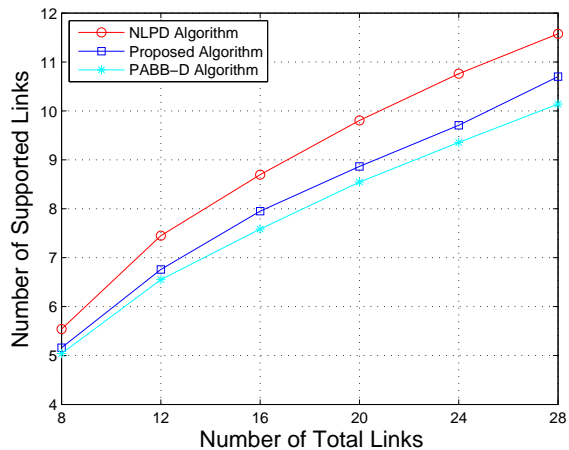}
\caption{Average number of supported links versus the number of total links.}\label{LNum}
\end{figure}

\begin{figure}[!t]
\centering\includegraphics[width=8.5cm]{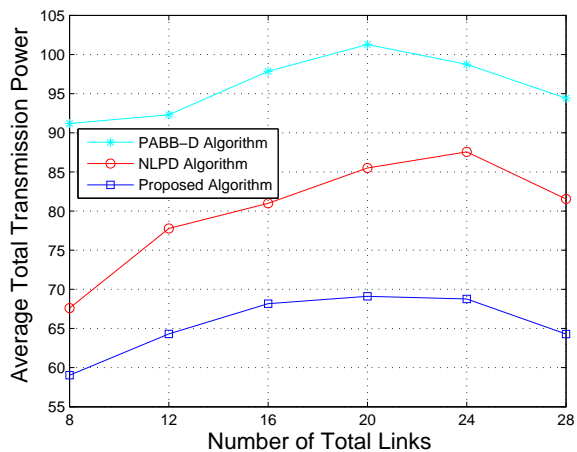}
\caption{Average total transmission power versus the number of total links.}\label{AverP}
\end{figure}

%

{Figs. \ref{LNum} and \ref{AverP} are obtained by averaging over $200$ Monte-Carlo runs.} They show that the proposed algorithm can support more links with significantly less total transmission power than the PABB-D algorithm in \cite{robust}. The performance improvement is due to the novel chance SINR constraint \eqref{chance2} and the JPAC formulation \eqref{model} based on it. Fig. \ref{AverP} shows that the NLPD algorithm in \cite{msp} transmits more power than the proposed algorithm. The reason for this is because the NLPD algorithm supports (much) more links than the proposed algorithm, as shown in Fig. \ref{LNum}. Notice that the NLPD algorithm in \cite{msp} requires the knowledge of perfect CSI and the set of supported links obtained by it (potentially) changes fast with channel fadings (and thus not robust). In contrast, our proposed algorithm is based on the CDI and can support a fixed set of links over a relatively long period of time (and thus robust). {One may say that it is not intuitive that the average total transmission power decreases with the  number of total links (when the number of links is relatively large) in Fig. \ref{AverP}. The reason why this ``strange'' phenomenon happens is because that as the number of total links increases, the diversity of links in the network increases and there is more degree of freedom to select a good subset of links to support (in terms that the links in the selected subset cause weak interferences with each other).}

{\begin{table}[!t]
\centering \caption{\textsc{Maximum outage probability among $200$ Monte-Carlo runs versus number of total links}}\label{passrate}
\begin{tabular}{|c| c c c c c c|}
\hline
 \tabincell{c}{Number of Total Links ($K$)} &  8 & 12 & 16 & 20 & 24 & 28\\
\hline
 \tabincell{c}{Outage Probability ($\times 10^{-3}$) } & 0.8 & 0.8 &1.2 &1.2 &1.2 &1.4 \\
\hline
\end{tabular}

\end{table}
}
{To check whether the SINR outage probability of the links in the supported set returned by the proposed algorithm is less than or equal to the preselected tolerance $\epsilon=0.05,$ we further generate $5000$ new samples according to the channel distribution for each $K$ and each Monte-Carlo run. Table I summarizes the maximum outage ratio (probablity) among $200$ runs. We can observe from Table I that the maximum outage ratio of all $200$ runs are (significantly) less than the given tolerance, which shows that the proposed algorithm can effectively guarantee the outage probability.}

 \end{document}